\documentclass[epj]{webofc}
\usepackage[varg]{txfonts}   
\usepackage{color}
\wocname{ARENA-2016}
\woctitle{ARENA-2016}
\begin{document}
\title{$X_\text{max}$ reconstruction from amplitude information with AERA}
%
%

\author{\firstname{Florian} \lastname{Gat\'{e}}\inst{1}\fnsep\thanks{\email{florian.gate@subatech.in2p3.fr}}
 \firstname{} \lastname{for the Pierre Auger Collaboration}\inst{2}}


\institute{SUBATECH - Ecole des Mines de Nantes - CNRS/IN2P3 - Universit\'{e} de Nantes
\and
            Full author list: http://auger.org/archive/authors\_2016\_06.html
          }

\abstract{%
The standard method to estimate the mass of a cosmic ray is the measurement of the atmospheric depth of the shower maximum ($X_\text{max}$). This depth is strongly correlated with the mass of the primary because it depends on the interaction cross section of the primary with the constituents of the atmosphere. Measuring the electric field, emitted by the secondary particles of an extensive air shower (EAS), with the Auger Engineering Radio Array (AERA) in the 30-80 MHz band allows the determination of the depth of shower maximum on the basis of the good understanding of the radio emission mechanisms. The duty cycle of radio detectors is close to 100\%, making possible the statistical determination of the cosmic-ray mass composition through the study of a large number of cosmic rays above 10$^{17}$ eV. In this contribution, $X_\text{max}$ reconstruction methods based on the study of the radio signal with AERA are detailed.
}
\maketitle

\section{Sensitivity of the radio emission to the nature of cosmic rays}
\label{sec-1}

\begin{figure}[h]
\centering
\sidecaption
\includegraphics[width=6cm,clip]{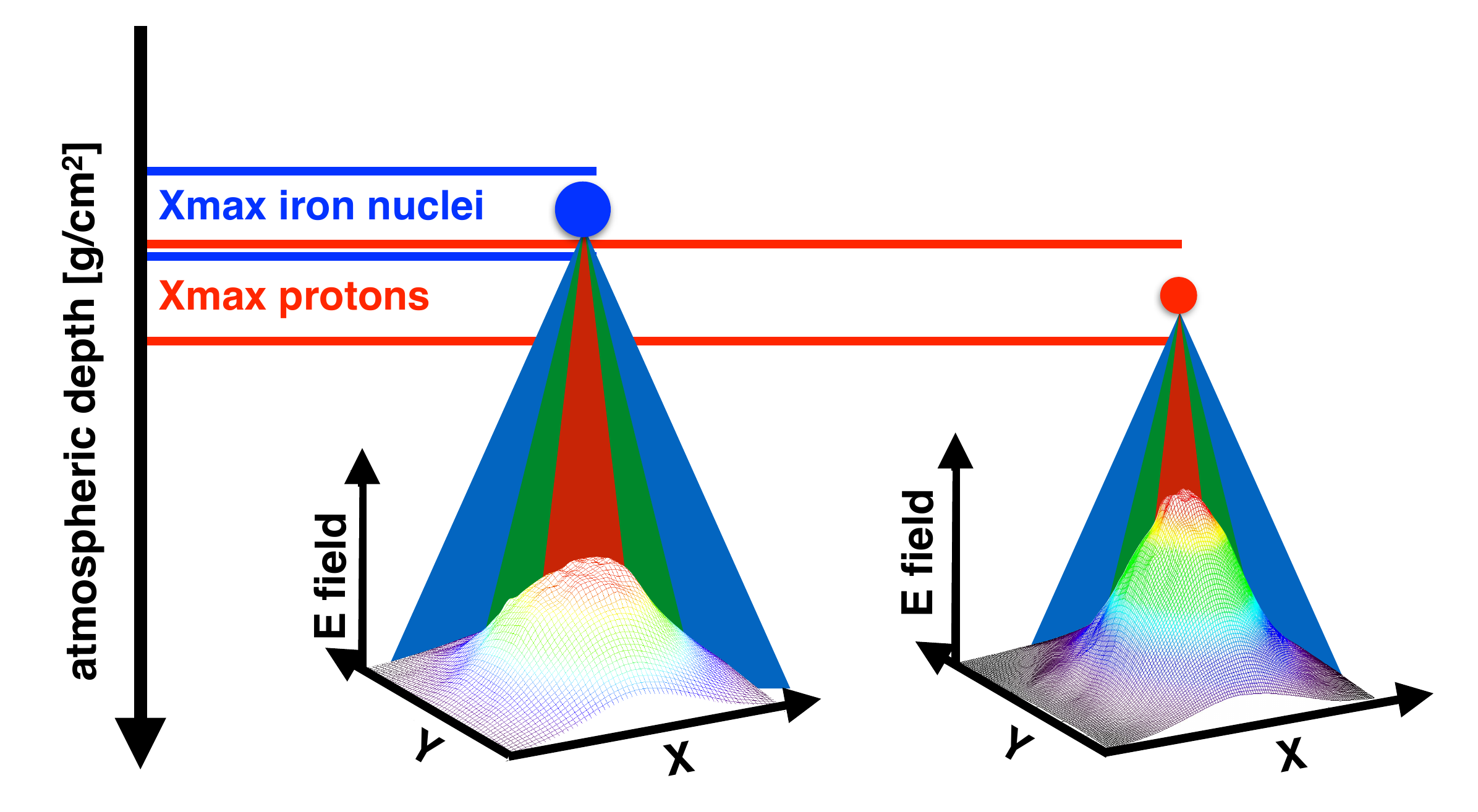}
\caption{Dependence of the foot print of the radio emission on the mass of the primary. Two LDF are simulated with SELFAS, on the left induced by an iron nucleus, on the right induced by a proton. The vertical axis represent the atmospheric depth and the horizontal lines account for the relative $X_\text{max}$ depth distributions for proton induced showers and iron nucleus induced showers.}
\label{arena1}       
\end{figure}

The AERA antennas \cite{AERA} record the electric field for two horizontal polarizations (North-South \& East-West directions). The arrival direction of the shower is calculated from the relative arrival time of the pulses at several stations. In the far-field approximation, the vertical component is calculated form the measured horizontal projection of the electric field and the arrival directions. The maximum signal received by an antenna is calculated as the maximum of the Hilbert envelope of the three polarizations. The radio emission is characterized by the lateral distribution function (LDF) of the received electric field. It can be calculated as the maximum electric field recorded by the triggered antennas, or the energy fluence, as a function of their position with respect to the shower axis. The topology of the electric field depends on the mass of the primary, as shown in Fig. \ref{arena1}.
Light nuclei are more likely to interact at lower altitude than heavier nuclei of the same energy. As the electric field emission is strongly beamed towards the direction of propagation of the shower, the LDF is narrower in the case of light nuclei compared to heavier nuclei. Thus, the radio signal is correlated to the mass information, that is estimated through the reconstruction of $X_\text{max}$, the atmospheric depth at which the number of secondary particles reaches its maximum. The features of the electric field that are known to be correlated to $X_\text{max}$ are the amplitude \cite{BuitinkLOFAR_Xmax2014, 2012ApelLOPES_MTD, TunkaRex_Xmax2016}, the shape of the radio wave front \cite{2014ApelLOPES_wavefront} and the spectral index of the frequency spectrum \cite{Grebe_ARENA2012}.

\vspace{-.2cm}
\section{Reconstruction methods}
\label{sec-2}

\begin{figure}[h]
\centering
\sidecaption
\includegraphics[width=6cm,clip]{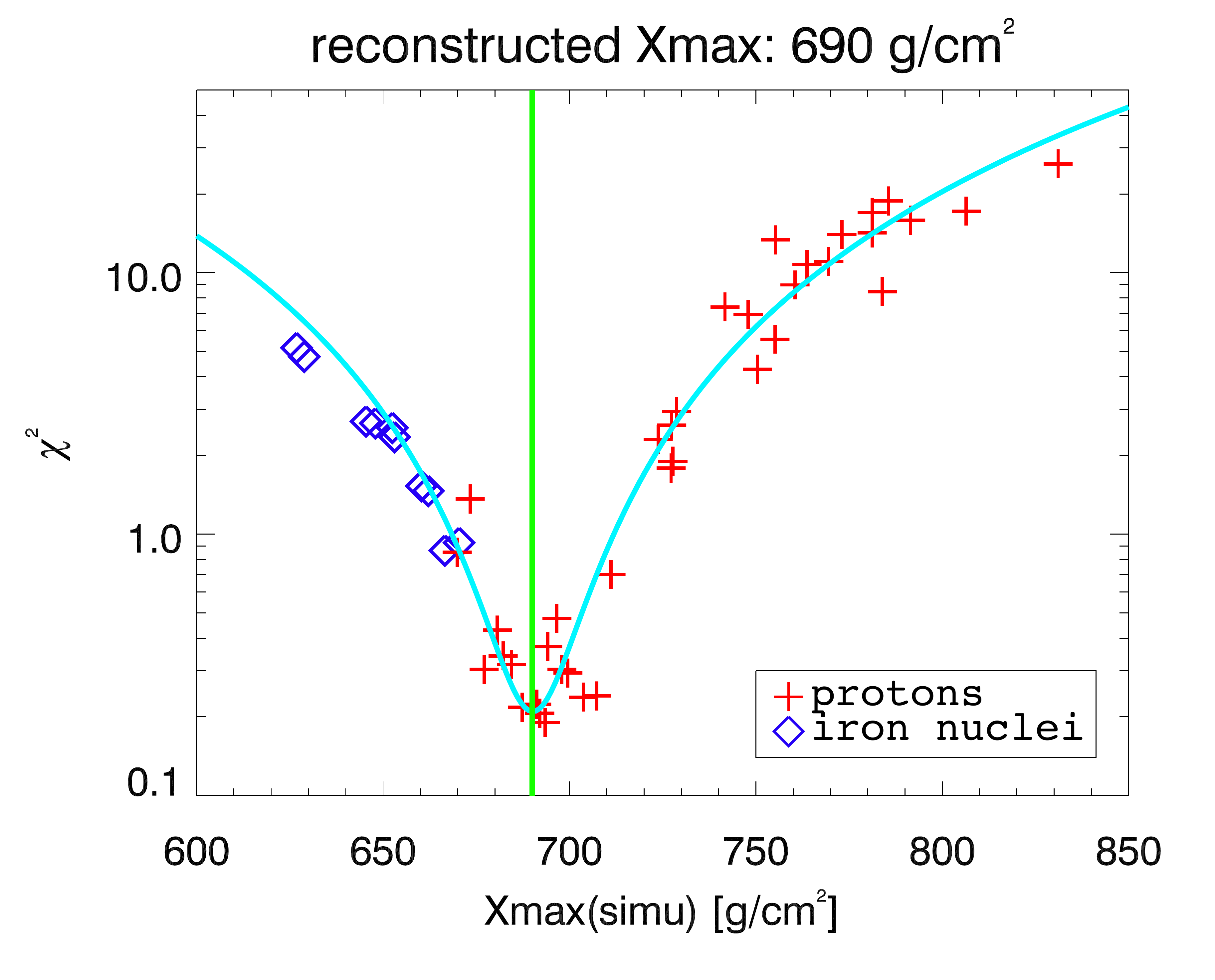}
\caption{Agreement between a detected LDF at AERA and the simulated LDFs as a function of their respective $X_\text{max}$ depths. The values are fitted by a square function represented by the curve. The $X_\text{max}$ depth reconstructed by the radio method for this event is given by the minimum of the square function (i.e. the value that gives the best agreement), highlighted on the plot by the vertical line.}
\label{arena2}       
\end{figure}

Four methods have been developed to perform a reconstruction of the depth of the shower maximum from AERA data. They are based on a comparison of the amplitude (or energy fluence) of the detected electric field to a prediction model. To reconstruct one detected shower, two sets of events are simulated with the experimentally reconstructed arrival direction, initiated by protons and iron nuclei. The total set is composed of a higher fraction of protons to account for the larger first interaction depth and $X_\text{max}$ fluctuations of the light nuclei. The LDFs corresponding to different $X_\text{max}$ values are computed by a simulation code and the agreement with the experimental data is tested as a function of the simulated $X_\text{max}$, with a $\chi^2$ test. The reconstructed $X_\text{max}$ is the minimum value of the function $\chi^2 = f(X_\text{max})$, as shown in Fig. \ref{arena2}. The methods (noted A, B, C, D) are detailed in the next section.

\vspace{-.1cm}

\vspace{-.5cm}

\textbf{A - SELFAS}: The electric field induced by 40 protons and 10 iron nuclei with an arbitral energy of 1 EeV is computed by SELFAS \cite{SELFAS} on a dense array. No assumption is made on the core position or $X_\text{max}$. The core position is estimated by shifting the simulated core position over the frame of AERA until the best agreement with the detected LDF is found. This procedure is done for all 50 simulations. A scaling parameter is added in the $\chi^2$ test in order to compare only the shapes of the LDFs and the electric field amplitude is assumed to be linearly proportional the energy of the primary.\\

\textbf{B - Parametrized model}: Instead of computing the electric field, this method uses a parametrized model of electric field emission in the shower frame \cite{2DLDF}. The $C_i$ parameters are fixed at values giving the best results for the AERA experiment from CoREAS simulations. The parameter $\sigma$ accounts for the width of the radio footprint on the ground. The parametrization allows the estimation of the geometrical distance to the shower maximum based on $\sigma$. Finally, knowing the arrival direction and an atmospheric model, the distance is converted to the corresponding crossed atmospheric depth $X_\text{max}$. 
\begin{equation}
u(\textbf{r}) = A \left [ \exp \left ( \frac{-(\textbf{r}+C_1 \textbf{e}_{\textbf{v}\times\textbf{B}} -\textbf{r}_\text{core})}{\textcolor{red}{\sigma}^2} \right ) - C_0 \exp \left ( \frac{-(\textbf{r}+C_2 \textbf{e}_{\textbf{v}\times\textbf{B}} -\textbf{r}_\text{core})}{(C_3 e^{C_4\textcolor{red}{\sigma}})^2} \right ) \right ]
\end{equation}

\textbf{C - CoREAS}: This method is similar to method A but, the electric field is computed by CoREAS \cite{CoREAS}. The primaries are simulated with the energy and arrival direction calculated from the data of the surface detectors (SD). The reconstruction is also based on the comparison of the simulated and measured radio-emitted energy, deposited on the ground by the secondary particles.\\

\textbf{D - ZHAireS}: The electric field is computed by ZHAireS \cite{ZHAireS} only along a line of antennas. It uses a superposition model \cite{superpos} to estimate the electric field around the shower axis from the line of antennas. Indeed, in the shower frame, the amplitudes of the geomagnetic and Askaryan contributions are circularly symmetric around the shower axis. The computing time is considerably reduced. This method also uses the SD energy as a prior.\\

\vspace{-.6cm}

\section{Comparison to fluorescence detection measurements}
\label{sec-3}

\begin{figure}[h!]
\centering
\includegraphics[width=6.5cm,clip]{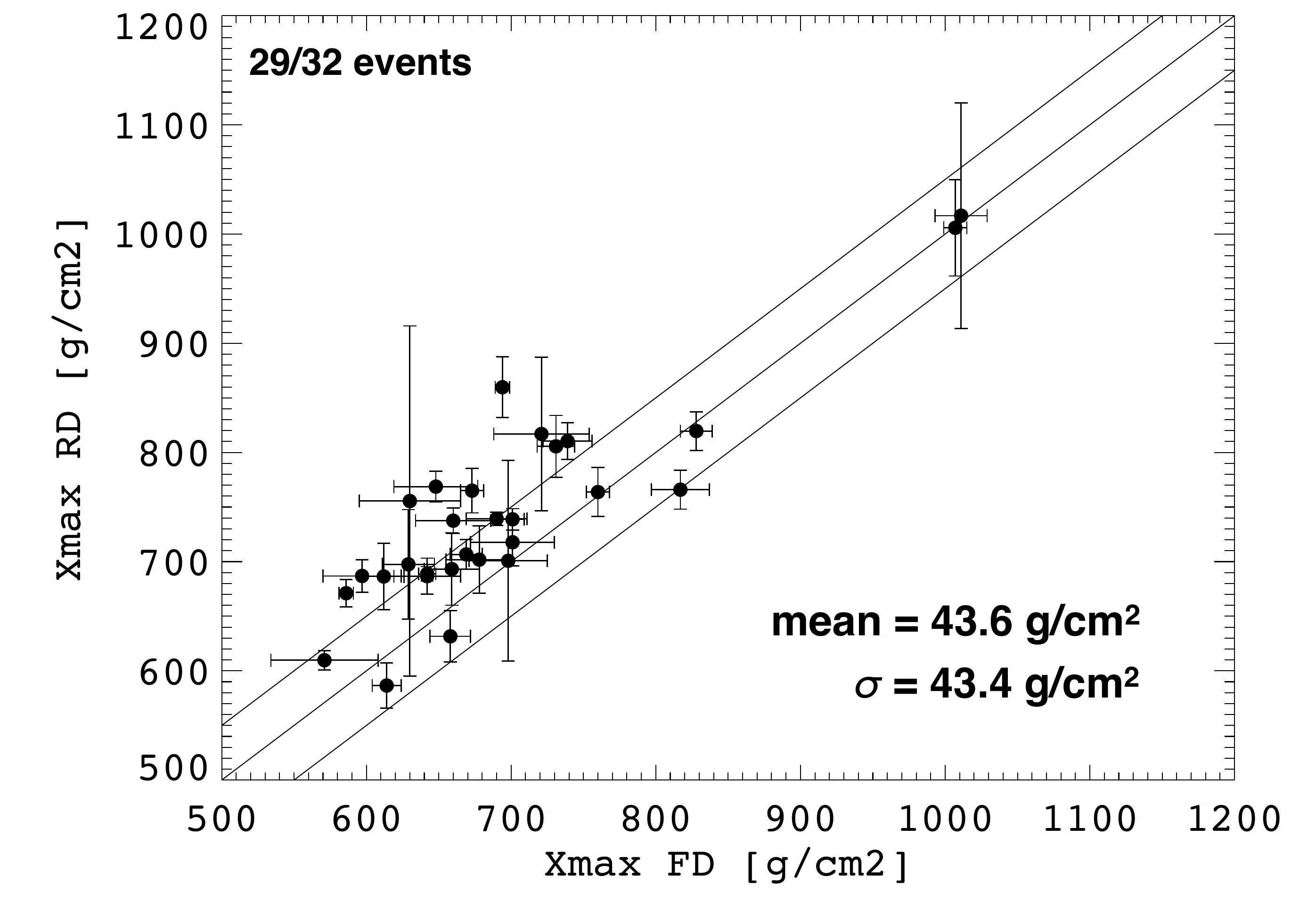}\hfill
\includegraphics[width=5.cm,clip]{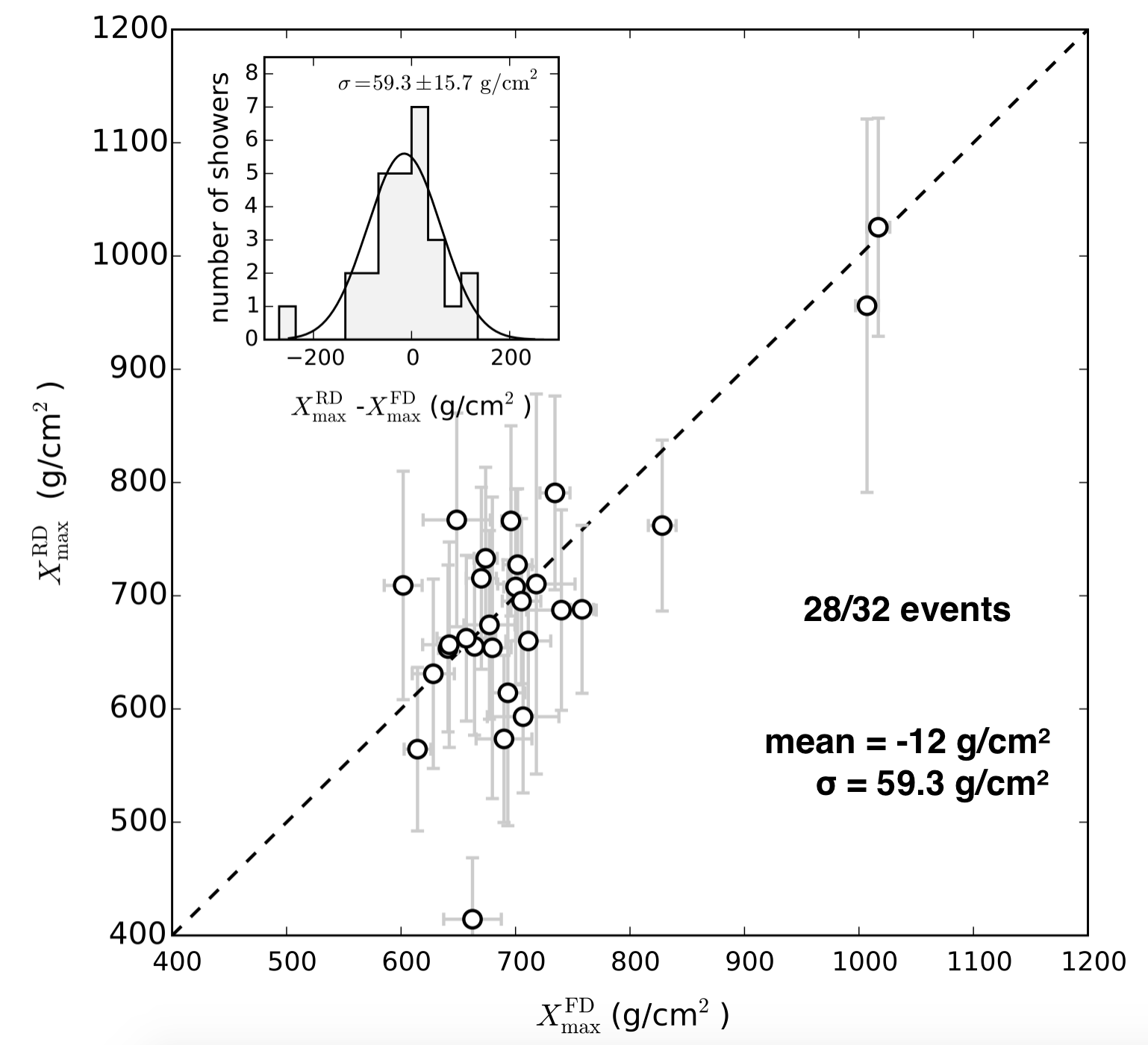}
\caption{Scatter plots obtained with method A (left) and method B (right). The reconstructed $X_\text{max}$ are displayed as a function of the FD measurements. The lines represent a one-to-one correlation, surrounded by two lines accounting for $\pm 50$ g/cm$^2$ on the left plot.}
\label{arena3}       
\end{figure}

\begin{figure}[h]
\centering
\includegraphics[width=6cm,clip]{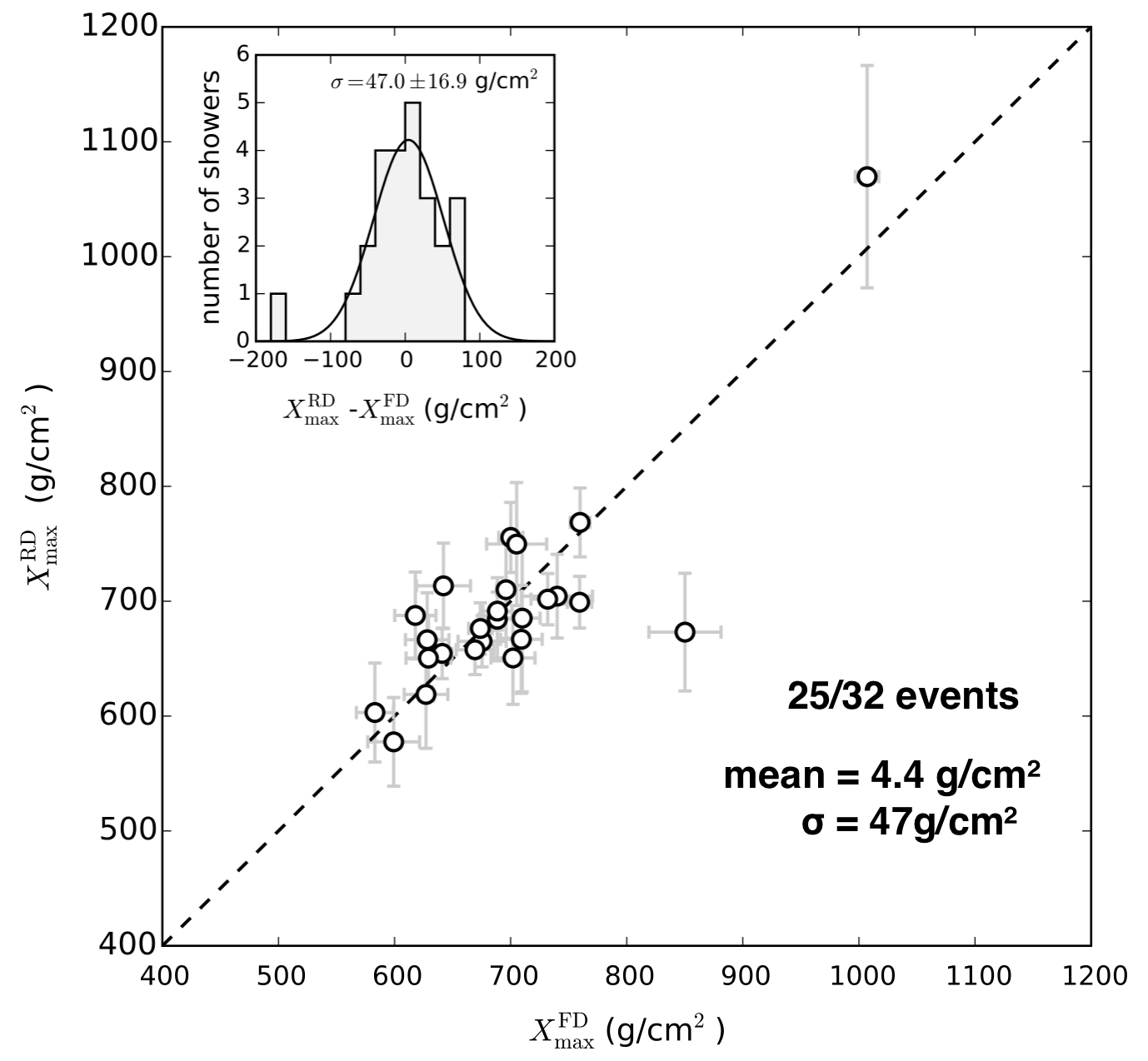}\hfill
\includegraphics[width=7.8cm,clip]{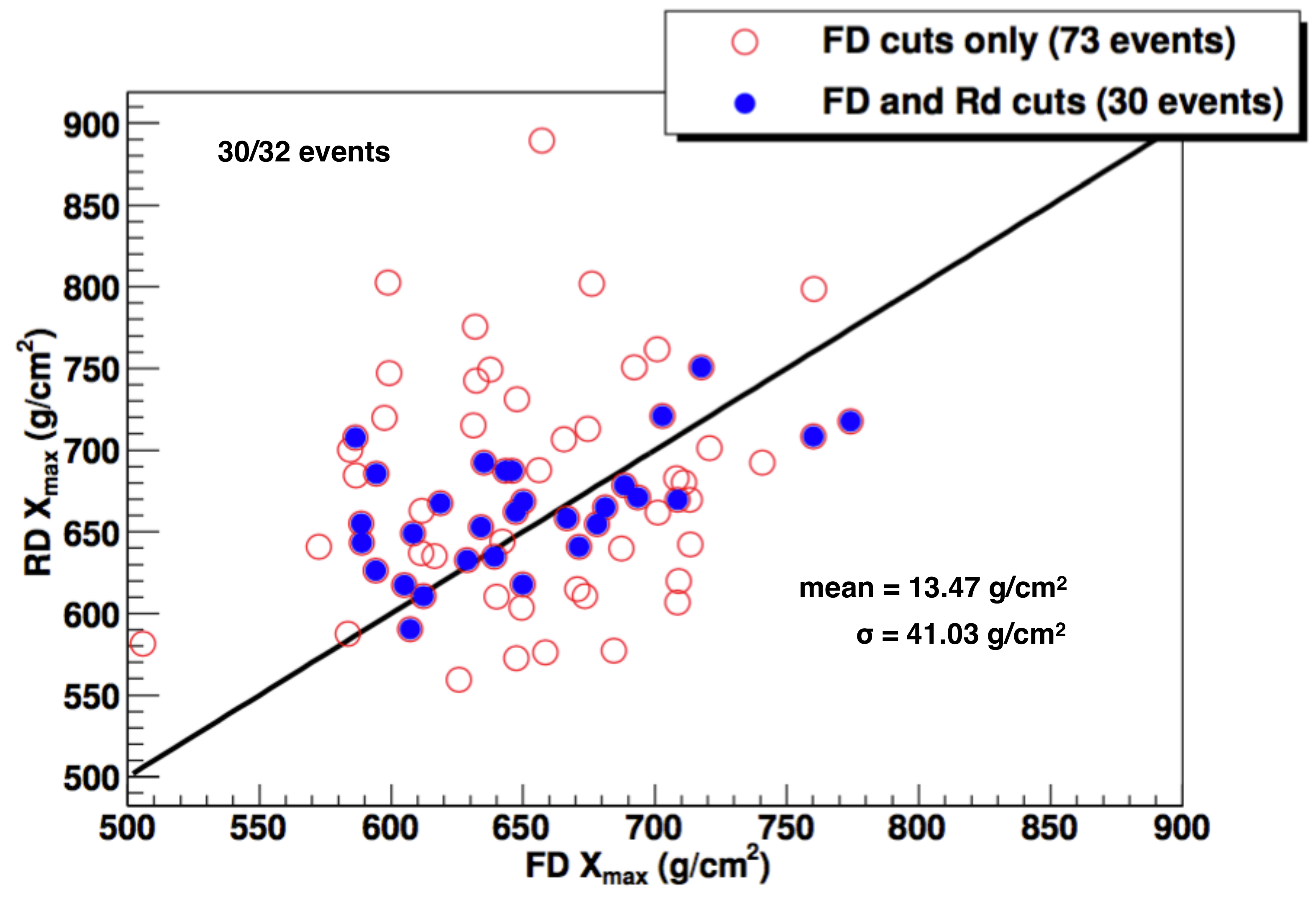}
\caption{Scatter plots obtained with method C (left) and method D (right).}
\label{arena4}       
\end{figure}

The methods described in Section \ref{sec-2} are applied on a high quality set of hybrid (radio + FD) showers. The official quality cuts of the fluorescence detector (FD) are applied, at least 5 radio stations must have trigged and the zenith angle must not exceed 55$^\circ$. The total number of events is 32. However, the number of events reconstructed by each method varies slightly. The combination of the event characteristics (energy of the primary, mass, $X_\text{max}$) may not be reproducible with the hadronic interaction model used by one of these methods. Multiple $X_\text{max}$ solutions can also be produced for a single event. The correlation plots between the reconstructed $X_\text{max}$ with the radio methods and the FD measurements are presented in Fig. \ref{arena3} and Fig. \ref{arena4}. The reconstructed values are in good agreement with the FD measurements. The maximum systematic deviation is of 40 g/cm$^2$ (method A).


\vspace{-1.cm}

\section{Conclusion and outlook}
\label{sec-4}

The distribution of the deviation of the reconstructed $X_\text{max}$ from the FD measurements for the set of hybrid showers ($\Delta X_\text{max} $) are fitted by a gaussian function and the mean difference and standard deviation at 1 $\sigma$ confidence level are summarized in Table \ref{tab-1} for each method. The offsets could be related to the underlying simulation codes. The reconstruction of $X_\text{max}$ from the radio signal will be a decisive asset for the estimation of the mass composition of cosmic rays due to the high duty cycle of the radio stations. The next steps are the determination of the composition from radio without any bias and the combination of the strength of the radio methods. To be efficient, the reconstruction needs, to be precise and with a fast calculation time. The validity of the methods can only be proven through the comparison with the FD measurements. The influence of the air density and refractivity at the time of the measurements on the deviations are under study. The $X_\text{max}$ resolution from radio methods is currently close to 40 g/cm$^2$.

\begin{table}[h]
\centering
\caption{Summary of the requirements and results of the radio methods}
\label{tab-1}       
\begin{tabular}{|c|c|c|c|c|}
\hline
method & A & B & C & D  \\\hline
requirements & RD direction & RD direction  & SD direction & SD direction  \\
 & &  & SD energy & SD energy \\\hline
calculation time (one event) & 8 hours & - - - - - - - - - -  & ~1 week & ~1 hour\\\
simulated antennas& 168  & 0 & 160 & 60 \\\
number of showers& 40 p + 10 Fe  & 0 & 20 p + 10 Fe & 30 p + 30 Fe \\\hline
mean($X_\text{max,RD} - X_\text{max,FD}$) (g/cm$^2$)& $43.6 $ & $-12 $ & $4.4 $ & $13.47$ \\\hline
$\sigma$ ($X_\text{max,RD} - X_\text{max,FD}$) (g/cm$^2$) &43.4 & 59.3 & 47  & 41.03 \\\hline

\end{tabular}

\end{table}

\vspace{-0.cm}

%

\begin{thebibliography}{}
%
%


\bibitem{AERA} J. Schulz for the Pierre Auger Collaboration, Proc. of the 34th ICRC, PoS(ICRC2015) 615, (2015)

\bibitem{BuitinkLOFAR_Xmax2014} S. Buitink et al. LOFAR Coll., Phys. Rev. D, (2014) 90


\bibitem{2012ApelLOPES_MTD} W.D. Apel, et al. LOPES Coll., Physical Review D (2012) 85, 10.1103/PhysRevD.85.071101

\bibitem{TunkaRex_Xmax2016} P.A. Bezyazeekov, et al. Tunka-Rex Coll., JCAP (2016) 01


\bibitem{2014ApelLOPES_wavefront}W.D. Apel, et al. LOPES Coll., JCAP (2014) 09


\bibitem{Grebe_ARENA2012} S. Grebe et al. Pierre Auger Coll., AIP Conference Proceedings, (2013) 1535


\bibitem{SELFAS} V. Marin, B. Revenu, Astropart. Phys. 35 (2012) 733-741

\bibitem{2DLDF}A Nelles et al, Astropart. Phys. 60 (2015) 13

\bibitem{CoREAS}Huege, T., Ludwig, M, James, C.W., AIP Conf. Proc. (2013) 128-132 

\bibitem{ZHAireS} J. Alvarez-Mu\~niz, W. R. Carvalho Jr., E. Zas, Astropart. Phys. 35 (2012) 325-34

\bibitem{superpos}J. Alvarez-Mu\~niz, W. R. Carvalho Jr., H. Schoorlemmer, E. ZasAstropart, Phys. 59 (2014) 29


\end{thebibliography}
%
%

\end{document}